\documentclass[prl,twocolumn,showpacs]{revtex4}
\usepackage{graphicx}

\begin{document}

\title[Command Option Summary]{Anomalies in heat capacity of YBa$_2$Cu$_3$O$_{6+x}$ in normal
state}

\author{V.N. Naumov$^1$, G.I. Frolova$^1$, V.V. Nogteva$^1$, A.I. Romanenko$^1$,
and T. Atake$^2$}%

\affiliation{$^1$Institute of Inorganic Chemistry, Siberian Branch
of Russian Academy of Science, Novosibirsk, Russia}
\affiliation{$^2$Tokyo Institute of Technology, Yokohama, 226,
Japan}

\date{\today}%

\begin{abstract}
The temperature dependence of heat capacity for the samples of 90
K phase of YBa$_2$Cu$_3$O$_{6+x}$ with $x$ = 0.85, 0.90 and 0.95
was investigated above $T_c$. For separating the heat capacity
into regular and anomalous contributions the special technique was
used. For all the samples the anomalies in intervals 100--200 K
($T_{low}$), 205--230 K ($T_m$) and 260--290 K ($T_h$) were
detected. At $T\simeq3T_c$ the temperature "echo" from
superconducting phase transition was discovered. This "echo" is
connected with the anomaly $T_h$, which is believed to mark the
rise of pairing carriers of charge without phase coherence. We
connect the anomaly $T_m$ with the Neel point of magnetic
subsystem which has antiferromagnetism coexisting with
superconductivity. The $x$ -- dependence of entropy $S_h$ points
out that $T_h$ -- process probably exists only in the frameworks
of 90 K phase. The curves $S_m$(x) and $S_h$(x) intersect each
other at the optimal doping. The discovered anomalies mark new
lines on the phase diagram of YBa$_2$Cu$_3$O$_{6+x}$ -- compounds.
\end{abstract}


\maketitle

In many properties of superconducting compounds
YBa$_2$Cu$_3$O$_{6+x}$ ($x>0.4$), besides anomalies resulted from
superconducting phase transition (at temperature $T_c$), anomalies
in the range of normal state ($T>T_c$) are observed (for example,
~\cite{1,2,3,4,5}).

The anomalies stably shown up in heat capacity and the
corresponding singularities in other physical properties of
YBa$_2$Cu$_3$O$_{6+x}$ inevitably raise the questions: to what
subsystem of superconductor (the lattice one, magnetic one or
electron one) - they are relevant, what processes they mark.
Perhaps these processes are caused by the rise of incoherent
pairing carriers of charge, by opening the pseudogap in the
spectrum of spin excitations, by manifestation of wave of charge
density, or by other phenomena \cite{6,7,8,9,10,11}. For clearing
up the noted questions the investigations of heat capacity of
YBa$_2$Cu$_3$O$_{6+x}$ in the range of the normal state seem to be
topical.

In this work we analyze the precise experimental data on heat
capacity of three sample of superconducting ceramics
YBa$_2$Cu$_3$O$_{6+x}$ ($x$ = 6.85, 6.90 and 6.95), which fall
into a region of 90 K phase ($T_c \simeq 92$ K). Our investigation
showed that three anomalies in the temperature interval 100--320 K
were observed. The obtained results indicate that anomalies in
heat capacity above $T_c$ mark the characteristic properties of
YBa$_2$Cu$_3$O$_{6+x}$ -- system rather than the imperfections of
individual sample.

The measurements of heat capacity $C_p(T)$ were carried out in
different laboratories by means of vacuum adiabatic calorimetry.
We used the experimental points presented in \cite{12,13,14}.

We set the experimental heat capacity $C_p(T)$ in the range above
$T_c$ to be presented by the expression
\begin{equation}
C_p(T)=C_v(T)+\gamma T +RA(T/T_0-1)^\alpha+\delta
C(T)\,,\label{first}
\end{equation}
where $R$ is the universal gas constant. Here the term $C_v(T)$
describes the harmonic part of lattice heat capacity; the term
$\gamma T$ consists in two components - the linear electron one
and the linear anharmonic one; the term of $RA(T/T_0-1)^\alpha$ is
used to approximate the low-temperature wing of the another
anomaly which is stably observed in the YBa$_2$Cu$_3$O$_{6+x}$--
system at the temperatures $T > 250$ K and which is caused by
modification of the oxygen subsystem in the plane of chains
CuO$_x$ \cite{12,15,16}; the term $\delta C(T)$ describes the
anomalous part of heat capacity (if it does exist). The harmonic
part of the lattice heat capacity $C_v(T)$ is presented by known
function of temperature with three parameters $\Theta _2$, $\Theta
_4$ and $\Theta _\ast$, which correspond to the second moment of
the phonon density of states, to the fourth moment and to the
effective moment attributed to the upper limit of a phonon
spectrum (the method of effective sum based on high temperature
expansion of heat capacity \cite{17,18,19}). For each investigated
sample the numerical values of parameters $\Theta _2$, $\Theta
_4$, $\Theta _\ast$ and $\gamma$ (entered the expression (1)) were
determined by least square method in the temperature interval
100--250 K. Then, two first terms were calculated and subtracted
from the experimental heat capacity $C_p(T)$ above 100 K.
Parameters of the term $RA(T/T_0-1)^\alpha$ were determined by
approximating the remainder in the interval 235--310 K (as an
example see \cite{12,15,16}). With parameters determined in such a
way we calculated the terms $C_v(T)$, $\gamma T$ and
$RA(T/T_0-1)^\alpha$ in the interval 100--350 K. It should be
noted that these terms are nonoscillating smooth functions. For
the sample $x$ = 0.85 the second term $\gamma T$ (at 200 K) is
less than 2\% of the total heat capacity and the third term
$RA(T/T_0-1)^\alpha$ does not exceed ~1.5\% of the total heat
capacity. For the others the orders of magnitude are the same.

In the temperature range of normal state the subtracting of smooth
contributions $C_v(T)$, $\gamma T$ and \mbox{$RA(T/T_0-1)^\alpha$}
from experimental heat capacity was carried out. As a result the
anomalies in the form of peaks

\begin{figure}[tbh]
\centerline{\resizebox{0.48\textwidth}{!}
{\includegraphics{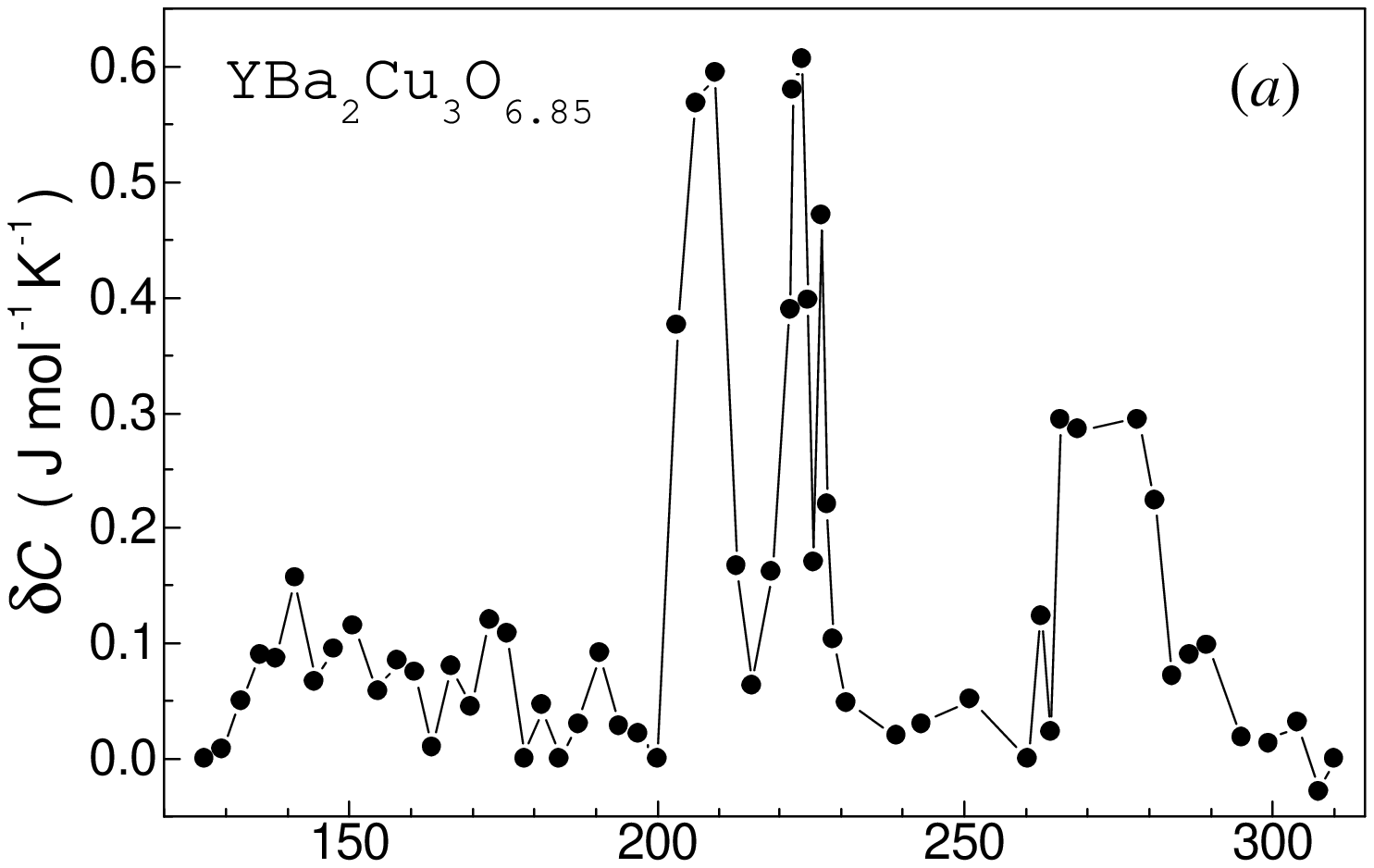}}}
\centerline{\resizebox{0.48\textwidth}{!}
{\includegraphics{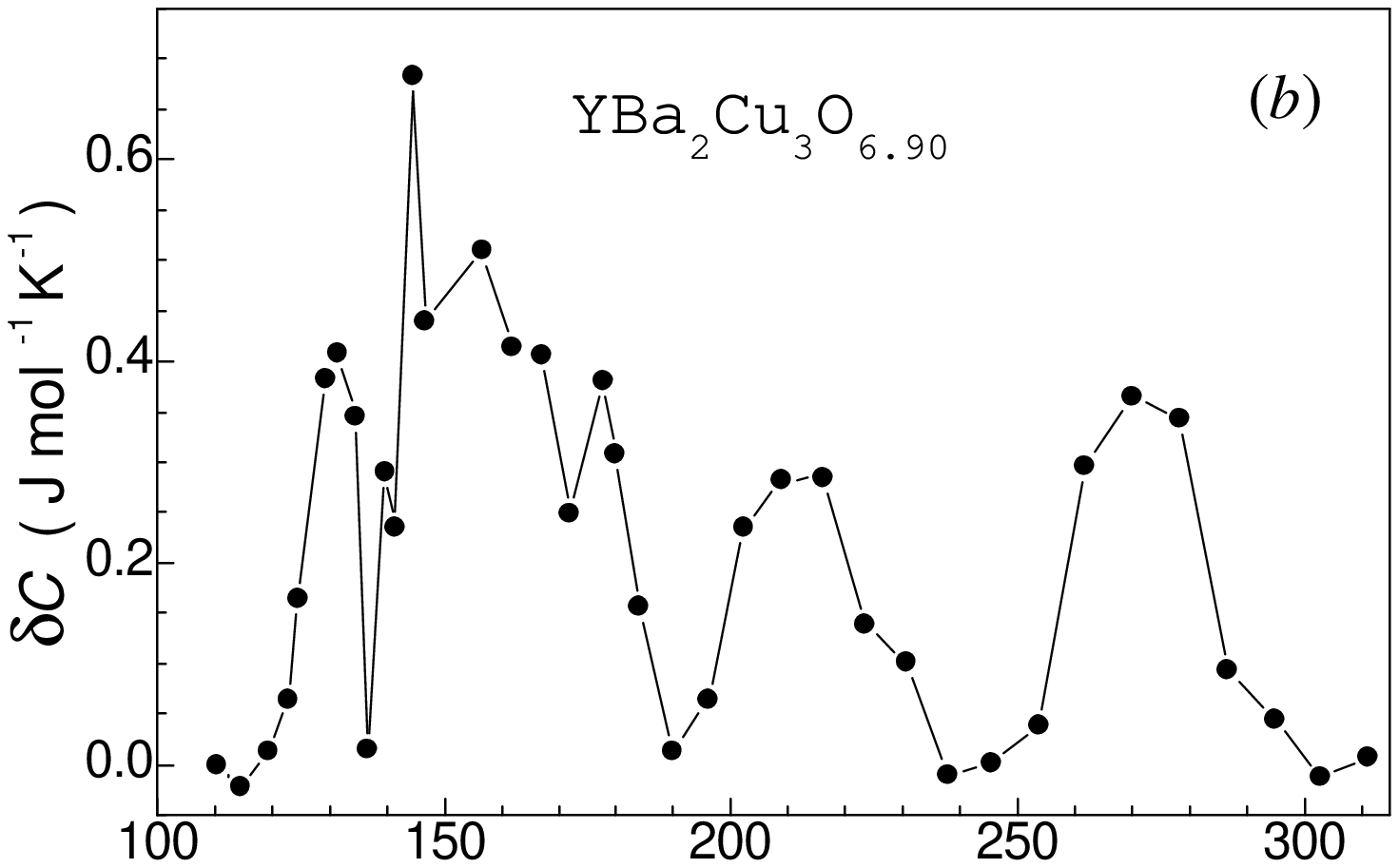}}}
\centerline{\resizebox{0.48\textwidth}{!}
{\includegraphics{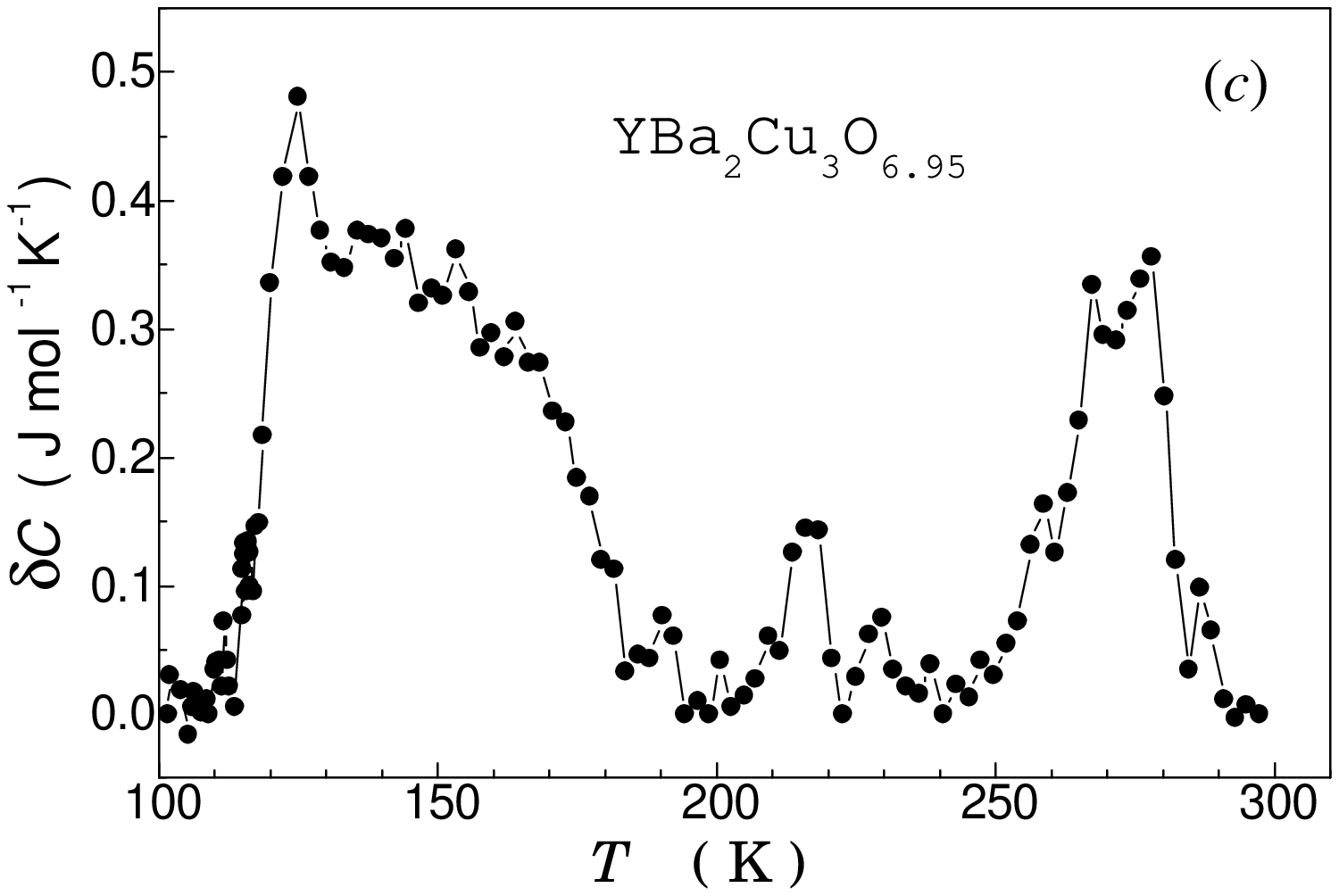}}} \caption[]{\label{fig1}Anomalies
$T_{low}$, $T_m$ and $T_h$ in heat capacity of the samples
YBa$_2$Cu$_3$O$_{6.85}$ (a), YBa$_2$Cu$_3$O$_{6.90}$ (b) and
YBa$_2$Cu$_3$O$_{6.95}$ (c)} \label{fig.1}
\end{figure}
higher then experimental scatter were discovered in the
temperature intervals 110--200 K ($T_{low}$), 205--230 K ($T_m$)
and \mbox{260--290 K} ($T_h$). The heights of peaks are different
for the samples with different $x$. These anomalies are presented
in {Fig.1}.

The anomaly $T_{low}$ occupies a wide temperature interval
(110--200 K). It would probably consist in several components. One
can see a separate peak in low-temperature region (110--140 K) for
all the samples. The anomaly $T_m$ has two peaks approximately of
the same height at 208 K and 225 K. The anomaly $T_h$ is located
in the range 260-290 K with a peak at $\simeq$ 275 K.

Further investigations should answer the questions to what
subsystem of superconductor -- to lattice one, magnetic one or
electron one -- these anomalies are relevant, and what their
nature is. In this work we consider the anomalies $T_h$ and $T_m$.

The analysis of experimental data shows that the correlation is
evident between the temperature $T_h$ and the temperature of
superconducting phase transition: $T_h\simeq 3T_c$ (276 K = 3 x 92
K). Thus the anomaly $T_h$ looks like a temperature "echo" arising
from superconducting phase transition at temperature $\simeq
3T_c$.

The temperature "echo" discovered in three investigated samples is
not the only example of such a phenomenon. Just the same
temperature "echo" from superconducting phase transition arising
at temperature $\simeq 3T_c$ was observed in the heat capacity of
the NdBa$_2$Cu$_3$O$_{6+x}$ and HoBa$_2$Cu$_3$O$_{6+x}$ compounds
as well \cite{15,20}.

The peculiarities in $T_h$ -- range are characteristic for
superconductors of YBa$_2$Cu$_3$O$_{6+x}$ system and they reveal
themselves in different properties. For example, for the sample
YBa$_2$Cu$_3$O$_{6.85}$ the derivative of electrical resistance
$\rho'(T)/\rho'(200 K)$ shows the sharp change just in the
temperature range 260--300 K (see Fig.4 in \cite{12}). Once more
example of the anomalous behavior of electrical resistance we
obtained for the sample YBa$_2$Cu$_3$O$_{6.9}$. The resistance
$\rho (T)$ was measured in the temperature interval 4.2--730 K
(Fig.2) by heating the sample in special regime \cite{21}. On the
curve $B$ in the range $T_h$, when the temperature decreases, the
character of conduction changes from the semiconducting one to the
metallic one. The deviation from semiconducting behavior begins at
$\simeq 275$ K. At this temperature some additional channel of the
conduction arises in the sample. Otherwise the semiconducting
behavior would continue further while temperature decreases. It is
possible this process represents the same phenomenon which one we
see in the heat capacity at temperature $T_h$.

\begin{figure}[tbh]
\centerline{\resizebox{0.36\textwidth}{!}
{\includegraphics{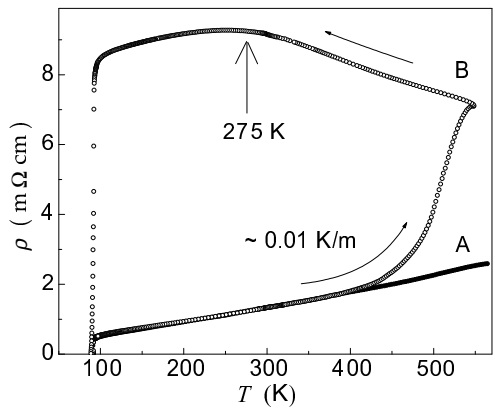}}} \caption[]{\label{fig2}Temperature
dependence of resistance: curve $A$ (black circles) --
conventional regime; curve $B$ (light circles) -- special regime
(see \cite{21})}
\end{figure}

One more example is presented in \cite{5} where the coefficient of
linear thermal expansion $\alpha (T)$ was measured along the three
orthorhombic axes for single crystals YBa$_2$Cu$_3$O$_{6.95}$ and
YBa$_2$Cu$_3$O$_7$. At temperature $T_c$ along each axis the
anomaly in $\alpha (T)$ was observed as a change of  slope.
Besides, for the sample YBa$_2$Cu$_3$O$_{6.95}$ the similar
anomalies were observed along each axis at temperature
\mbox{$\simeq 280$ K} (in our denotes $T_h$). One can consider
these anomalies, as an additional evidence of the temperature
"echo" from superconducting phase transition $(280 K : 3 = 93.3
K)$. It should be noted that anomalies in $\alpha (T)$ at
temperature $T_h$ are present only in the sample $x$ = 0.95,
whereas in the sample $x$ = 1 they are absent. This suggest that
$T_h$-- process depend on the doping.

The connection between temperature $T_h$ and temperature of
superconducting phase transition allows one to suppose that this
anomaly reveals some process concerned with the rise of
superconductivity, for example, with forming the Cooper pairs
above the temperature of superconducting phase transition \cite{6}
or with some other processes, from where the pseudogap in the
single-particle spectrum comes [22].

The anomaly in heat capacity, as a peak at temperature $T_h$,
evidences that this process happens as true phase transition
rather than a crossover. This conclusion is in agreement with the
view given in \cite{23} that transition to pseudogap phase is
accompanied by some hidden break of symmetry.

The small amplitude of the observed anomaly results from
microscopic size of the domains where the $T_h$ -- process happen.
However, the great number of these domains in the sample (as many
as the Avogadro number) gives the chance to observe it.

We assign a following meaning to introduced notion of the
temperature "echo". At relatively high temperature the new phase
arises in the short-range order which manifests itself by a phase
transition ($T_h$ -- process). Yet another phase transition (when
temperature decreases) is connected with development of the
long-range order ($T_c$ -- process).

Anomaly $T_m$ (with two peaks at $\simeq 210$ K and $\simeq 230$
K) can be observed not only in our compounds
YBa$_2$Cu$_3$O$_{6+x}$ and in (R)Ba$_2$Cu$_3$O$_{6+x}$
\cite{15,20}, but one can note the similar anomaly in heat
capacity of La$_{2-x}$Sr$_x$CuO$_4$ -- compound \cite{24}).

The temperatures of these two peaks almost coincide with the
points of magnetic phase transitions in pure CuO: 212 K and 230 K
\cite{25,26}. The estimation carried out for the sample $x$ = 0.85
shows that this anomaly can't be attributed to the impurity of
pure CuO, otherwise it would demand $\simeq 20$ mol $\%$ of
impurity to supply the observed contribution to heat capacity,
which absolutely disagrees with characteristic of the sample.

We dare to say that to account for anomaly $T_m$ in cuprate
systems, one can use the available idea of coexisting the
antiferromagnetism with superconductivity.
\begin{figure}[tbh]
\centerline{\resizebox{0.36\textwidth}{!}
{\includegraphics{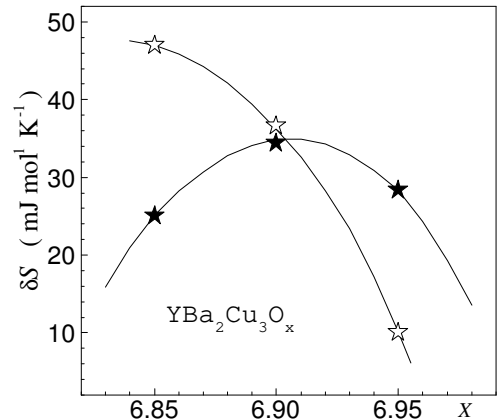}}} \caption[]{\label{fig3} $X$--
dependence of entropy: anomaly $T_m$ (light stars) and anomaly
$T_h$ (black stars)} \label{fig.3}
\end{figure}
The presence of two kind of antiferromagnetism was discussed
theoretically in \cite{23} and also some experimental evidence
were presented in \cite{27,28,29,30}. The presence of
antiferromagnetism coexisting with superconductivity was revealed
by elastic neutron scattering in superconductors of the family
La$_2$CuO$_4$ \cite{27,28} and also in superconductors
YBa$_2$Cu$_3$O$_{6.5}$ and
YBa$_2$(Cu$_{1-y}$Co$_y$)$_3$O$_{7+\delta}$ \cite{29,30}. It is
possible that anomaly $T_m$ marks the phase transition in this
magnetic subsystem (the Neel point).

In Fig.3 the $x$ -- dependence of entropy of $T_m$ -- and $T_h$ --
processes is presented. The dependence $S_h(x)$ is a dome with
maximum just in the range of optimal doping ($x\simeq 0.90$).

When $x$ decreases below the optimal doping range the entropy
$S_h(x)$ decreases as well. It is possible that it reduces to zero
at the lower boundary of 90 K phase of YBa$_2$Cu$_3$O$_{6+x}$--
compound. When $x$ increases above the optimal doping range the
entropy $S_h(x)$ decreases again. It looks as if $S_h(x)$ reaches
the zero value at $x$ = 7. The absence of anomaly $T_h$ in the
samples YBa$_2$Cu$_3$O$_{6+x}$ with $x$ = 7 is also confirmed by
the data of $\alpha (T)$ -- dependence \cite{5} (see above). Thus,
the $T_h$ -- process seems to accompany the superconductivity of
90 K phase of YBa$_2$Cu$_3$O$_{6+x}$-- compounds.

The entropy $S_m(x)$ sharply decreases when x increases above the
optimal doping. It is possible that it achieves its ultimate zero
value (disappearance of anomaly $T_m$) in so called overdoped
range, somewhere at $x$ less then 7. When x decreases in the range
lower than 6.85 the entropy $S_m(x)$ increases yet, but its rise
slows down now. It is possible that dependence $S_m(x)$ passes
through a maximum at some value $x$ and then it decreases,
reaching the zero value at the lower boundary of superconducting
phase ($x\simeq$ 6.4).

The curve $S_m(x)$ intersects the curve $S_h(x)$ in the point of
optimal doping. In this point the entropies of $T_m$-- and $T_h$--
processes are equalized. It is possible that this condition just
determines the point of optimal doping.

In summary, the accurate description of regular contributions
allows us to reveal the anomalies in temperature intervals
$T_{low}$ (110--200 K), $T_m$ (205--230 K) and $T_h$ (260--290 K)
for all the samples YBa$_2$Cu$_3$O$_{6+x}$ ($x$ = 0.85, 0.90 and
0.95).

It has been discovered that between the temperature $T_h$ and the
temperature of superconducting phase transition $T_c$ the
correlation takes place $T_h\simeq3T_c$. Thus the anomaly $T_h$
can be considered as a temperature "echo" from superconducting
phase transition. The analysis of data has shown that such a
temperature "echo" is observed for the samples
YBa$_2$Cu$_3$O$_{6+x}$ and (R)Ba$_2$Cu$_3$O$_{6+x}$, it being
observed not only in heat capacity but also in other properties.
The anomaly $T_h$ can be believed to reflect some process which is
connected with the arise of superconductivity.

For the explanation of anomaly $T_m$ the idea was used about the
existence in the cuprate superconductors of the second
antiferromagnetic subsystem, this antiferromagnetism coexisting
with the superconductivity. The anomaly $T_m$ (230 K) is believed
to mark the phase transition (the Neel point) in this magnetic
subsystem.

The dependence $S_h(x)$ of anomaly $T_h$ shows that $T_h$ --
process accompanies the superconductivity of 90 K phase of
YBa$_2$Cu$_3$O$_{6+x}$ compounds: its entropy has a maximum in the
range of optimal doping and possibly disappears on the boundaries
of this phase. The curves $S_h(x)$ and $S_m(x)$ intersect each
other in the point of optimal doping. It is possible that just
this condition determines the point of optimal doping of
superconductor.

It should be noted that the discovered anomalies mark new lines on
the phase diagram of YBa$_2$Cu$_3$O$_{6+x}$ -- compounds. \\


\begin{thebibliography}{99}
\bibitem{21}
  When heating and cooling the sample with the conventional for
  such experiments rate ($\sim10$ K/min), one ordinarily obtains the
  curve $A$. The curves of the resistance, on heating and cooling with
  such a rate, coincide. The curve $B$ can be obtained only in using
  the special temperature regime: the heating with the rate $\sim10$
  K/min up to \mbox{350 K} and then the deceleration of the heating rate
  down to $\sim 0.01$ K/min. Then on the rapid cooling (with the rate
  $\sim 10$ K/min) the dependence $\rho(T)$ follows another way (see Fig.2).
  Farther, on heating and cooling with the rate $\sim10$ K/min the
  resistance of the sample remains on the high-resistence branch
  of the curve $B$. (The reasons of such a behavior of the resistance will
  be discussed in the separate work).
\end{thebibliography}
\end{document}